\documentclass[10pt]{article}
\usepackage{amssymb}
\usepackage{amscd}
\usepackage{latexsym}
\usepackage{boldtensors}
\usepackage{amsmath}
\usepackage{amsthm}
\usepackage{graphicx}
\newcommand{\Zz}{{\mathbb Z}}

\newcommand{\Rr}{{\mathbb R}}

\def\be{\begin{equation}}
\def\ee{\end{equation}}
\def\bea{\begin{eqnarray}}
\def\eea{\end{eqnarray}}

\def\d{{\,\rm d}}
\def\i{{\,\rm i}}

\def\0{{\bf 0}}

\def\p{{\bf p}}

\def\h2m{\frac{\hbar^2}{2m}}
\def\p0{{P_{\beta H^0_N}}}

\theoremstyle{remark}  
\begin{document}
\title{%{\flushleft{\small {\rm Published in }\\}}\vspace{1cm}
\large\bf Momentum coupling of classical atoms}
\author{Andr\'as S\"ut\H o
\\Wigner Research Centre for Physics,  Hungarian Research Network\\
Email: suto.andras@wigner.hu}
\date{}
\maketitle
\thispagestyle{empty}
\begin{abstract}
\noindent
A new method, dual-space cluster expansion, is proposed to study classical phases transitions in the continuum. It relies on replacing the particle positions as integration variables by the momenta of the relative displacements of particle pairs. Due to the requirement that the particles must be static, coupling via the momenta partitions the set of particles into a set of clusters, and transforms the partition function into a sum over the different cluster decompositions. This allows us to derive a formula for the density that finite clusters can carry in the infinite system. In a simplified example, we then demonstrate that in two and higher dimensions this density has a threshold, beyond which the particles form infinite clusters. The transition is accompanied by a singularity in the free energy. We also show that infinite clusters are always present in condensed phases, most likely submacroscopic in liquids and macroscopic in crystals.
\end{abstract}

\section{Introduction}

Proving phase transitions from first principles for classical particles in continuous space is perhaps the oldest persisting problem in statistical mechanics. For a rigorous proof of the vapor-liquid and the liquid-solid transitions in three dimensions for a realistic pair interaction, such as the Lennard-Jones potential, a major breakthrough would still be needed. Similar iconic unsolved problems are the fluid-solid transition in a three-dimensional system of hard spheres and the transition to a hexatic phase in two dimensions.
Past research has focused primarily on two questions: the van der Waals theory of condensation and phase separation in the Widom-Rowlinson model.

The derivation of van der Waals' equation of state was first realized in one dimension. Previously, Kac~\cite{K} had shown how to calculate the grand partition function for a one-dimensional gas with a potential constructed from a hard core and attraction between particles at $x_i$ and $x_j$ of the form $-\alpha e^{-\gamma|x_i-x_j|}$. In this model there is no phase transition, however, by replacing $\alpha$ with $\alpha_0\gamma$ and sending $\gamma$ to zero after the thermodynamic limit, van der Waals' equation of state as well as Maxwell's equal-area rule could be reproduced~\cite{KUH}. The method applied was dimension-specific, the same result in higher dimensions required new ideas, partly originating from van Kempen~\cite{vK}, and was obtained by Lebowitz and Penrose~\cite{LP}.

The Widom-Rowlinson model~\cite{WR} consists of a two-component gas in which like particles do not interact and there is a hard-core repulsion between particles of different types. First Ruelle~\cite{R1} proved phase separation at high activities, then Lebowitz and Lieb~\cite{LL} extended the result to finite-range positive pair potentials that drop discontinuously to zero at their range. Later the result was further extended to multi-component systems with more general interactions~\cite{DH1, DH2, GH}.

A major advance on vapor-liquid coexistence was made in the nineties by Johansson~\cite{J} and by Lebowitz, Mazel and Presutti~\cite{LMP}. Johansson worked on one dimensional systems interacting via Lennard-Jones type pair potentials, and proved separation into a low- and a high-density phase at low temperatures. Lebowitz {\em et al.} demonstrated vapor-liquid transition in $d\geq 2$ dimensions for Kac type interactions with four-body repulsive and two-body attractive forces.
Concerning the fluid-solid transition the results are extremely meager, we can mention only two toy models. One is a one-dimensional system which exhibits normal thermodynamic behavior at high temperatures and freezes into a prescribed configuration at low temperatures~\cite{S1}. The other is a two-dimensional model of "zipper molecules" which displays fluid-solid transition~\cite{BLRW}.

In this paper we propose a new approach to the problem of phase transitions in systems of classical particles in the continuum. It can be called the dual-space cluster expansion. The interaction, which we restrict to translation-invariant pair potentials, acts by correlating the motions of the particles. For interactions that decay sufficiently rapidly with the distance the correlation is weak at low density. By raising the density a qualitative change in the correlated motion of atoms can take place. The position and the momentum of classical particles are independent, the partition function factors accordingly, and any phase transition can arise only from the position-dependent configurational partition function, an integral of the $N$-particle Boltzmann factor over the different configurations of $N$ motionless particles in a bounded volume. By Fourier-expanding the two-particle Boltzmann factors and integrating with respect to the positions one can introduce a new set of variables, the momenta of the relative displacements of particle pairs. For $N$ position vectors there are $N(N-1)/2$ momenta, so they are subject to constraints, the physical meaning of which is to ensure that the new variables still describe a system of immobile particles. Analysis of the constraints reveals that they are of two types, graphical and numerical. A set of momenta can be represented on a graph whose $N$ labeled vertices indicate the particles and whose edges correspond to the nonzero momenta. A properly directed momentum vector is assigned to every edge, and at each vertex the sum of the incident momenta -- the total momentum of the corresponding particle -- must be zero. A graph on which this condition can be satisfied is called valid. We will see that the valid graphs are cyclic, that is, each edge is in at least one cycle. The graph is decomposed into maximal connected subgraphs, called clusters. Our gain from the Fourier expansion is precisely this decomposition, allowing us to derive a formula for the density of particles in finite clusters in the infinite system. The question then is whether, at low density, there are only finite clusters, and whether, as the density increases, infinite clusters appear. Such a transition is reminiscent of percolation in lattice models, but there could be more than one infinite cluster, and the singularity at the transition point would be stronger.

We will show that indeed, at low density there are only finite clusters. The appearance of infinite clusters depends on the convergence of an activity expansion {\em at} the convergence radius. The difficulty to prove it is partly combinatorial, we would need a
%precise asymptotic estimate of the number
good knowledge of the statistics of connected cyclic graphs on $n$ labeled vertices.
An additional difficulty is that the coefficients of the activity expansion are obtained as a sum over these graphs, and interaction enters the sum in a complex way. The best we can do is to preserve only the $(n-1)!/2$ cycles of length $n$. With this approximation, for a repulsive pair potential decaying as a Gaussian at large distances and logarithmically diverging at the origin, for dimensions two and beyond we find that infinite clusters appear at high densities. Furthermore, without any approximation we can still prove that condensed phases contain infinite clusters. The vapor-liquid transition can be associated with the appearance of infinitely many infinite but submacroscopic clusters; on the other hand, crystallization is expected to be accompanied by the formation of one or more macroscopic clusters.

In Section 2 we derive a recurrence relation, expressing $Q_{N,L}$, the canonical partition function for $N$ particles on the torus of side $L$, as a linear combination of $Q_{N-n,L}$, $n=1,\dots,N$. By dividing with $Q_{N,L}$ the resulting equation must be valid in the thermodynamic limit. Assuming that there are no infinite clusters, the limiting equation will be obtained in Section 3. In Section 4 we discuss the simplified example described above, and show that the equation derived in Section 3 fails at sufficiently high densities, so infinite clusters must appear. Section 5 presents the proof that condensed phases necessarily carry infinite clusters. The paper ends with a brief Summary.

\section{Recurrence formula for the partition function}\label{recurrence}

In this study the interactions are translation invariant even pair potentials: if $x_i, x_j\in\Rr^d$, then
\[
\tilde{u}(x_i,x_j)=u(x_i-x_j)=u(x_j-x_i).
\]
$u$ is real, satisfies the stability condition [R2], it is bounded below and tempered: there exist $\eta>d$ and $R,c>0$ such that
\[
|u(x)|\leq c|x|^{-\eta}\quad\mbox{if}\quad |x|>R.
\]
As a consequence, for any $\beta\in[0,\infty)$ the function $v(x)=1-e^{-\beta u(x)}$ is bounded and integrable, and $v\in L^1(\Rr^d)$ implies that  its Fourier transform
\[
 \hat v(\kappa)=\int e^{-\i 2\pi\kappa \cdot x}v(x)\d x
\]
exists and is continuous. We will need more, $\hat v\in L^1(\Rr^d)$, so both $v$ and $\hat v$ are continuous. $-v$ is the central object in the cluster expansion according to Mayer~\cite{M}. Here its Fourier transform will play the key role.
%{\bf  Note that $\hat v$ depends on $\beta$. If $u\geq 0$, then $\hat v(0)>0$ for all $\beta>0$. If $u$ has a negative part, by stability $\hat v(0)$ is positive for small $\beta$, but becomes negative and tends to minus infinity as $\beta$ tends to infinity. The most important example of this is the Lennard-Jones potential.}

The complete canonical partition function for $N$ identical particles on the hypercube $\Lambda=[-L/2,L/2]^d$ with periodic boundary conditions is
\be\label{partition-function}
Q_{N,L}=\frac{1}{\lambda_\beta^{dN} N!}\int_{\Lambda^N}\d x_1\dots\d x_N\prod_{1\leq i<j\leq N}E_L(x_j-x_i).
\ee
In this formula $\lambda_\beta=\sqrt{2\pi\hbar^2\beta/m}$ is the thermal wavelength for particles of mass $m$, $\lambda_\beta^{-dN}$ is the kinetic, the rest is the configurational partition function, and
\[
E_L(x_j-x_i)=e^{-\beta u_L(x_j-x_i)},\quad u_L(x)=\sum_{z\in\Zz^d}u(x+Lz).
\]
Because of the periodic boundary condition, in finite volumes we must replace $u$ with the periodic $u_L$.
From our assumptions on $u$ it follows that the lattice sum is absolutely convergent for $|x|>R$. Moreover, for any fixed $x$, $u_L(x)\to u(x)$ and
\[
v_L(x)=1-E_L(x)=1-e^{-\beta u_L(x)}\to 1-e^{-\beta u(x)}=1-E(x)=v(x)
\]
at rate $L^{-\eta}$ as $L\to\infty$. Note that $u_L(y)=u_L(-y)$ and $E_L(y)=E_L(-y)$.

The Boltzmann factors are Fourier-expanded,
\be\label{E-expanded}
E_L(x_j-x_i)=\sum_{z^j_i\in\Zz^d}\hat E_L(z^j_i)e^{\i \frac{2\pi}{L}z^j_i\cdot (x_j-x_i)},
\quad
\hat E_L(z)=L^{-d}\int_\Lambda e^{-\i \frac{2\pi}{L}z\cdot y}E_L(y)\d y.
\ee
$2\pi\hbar z^j_i/L$ is the momentum conjugate to $x_j-x_i$, where $i<j$; $\hat E_L(z)=\hat E_L(-z)$ is real. Substituting for $E_L(x_j-x_i)$ its  expansion (\ref{E-expanded}),
\begin{eqnarray*}
\prod_{1\leq i<j\leq N}E_L(x_j-x_i)=
%\prod_{1\leq i<j\leq N}\sum_{z^j_i\in\Zz^d}\hat E_L(z^j_i)\prod_{k=1}^Ne^{\i \frac{2\pi}{L}Z_k\cdot x_k},
\sum_{\{z^j_i\in\Zz^d|1\leq i<j\leq N\}}\left(\prod_{1\leq i<j\leq N}\hat E_L(z^j_i)\right)\prod_{k=1}^Ne^{\i \frac{2\pi}{L}Z_k\cdot x_k},
%\quad\mbox{\bf r.h.s. changed}
\end{eqnarray*}
where
\[
Z_k=\sum_{i=1}^{k-1}z^k_i-\sum_{j=k+1}^N z^j_k.
\]
$2\pi\hbar Z_k/L$ is the momentum of the $k$th particle. Integration with respect to $x_k$ constrains it to zero, yielding
\bea\label{partition-1}
Q_{N,L}
&=&\frac{L^{dN}}{\lambda_\beta^{dN}N!}\sum_{\{z^j_i\in\Zz^d|1\leq i<j\leq N\}}\left(\prod_{1\leq i<j\leq N}\hat E_L(z^j_i)\right)\prod_{k=1}^N \delta_{Z_k,0}
\nonumber\\
&=&\frac{L^{dN}}{\lambda_\beta^{dN}N!}\prod_{1\leq i<j\leq N}\left(\hat E_L(0)+\sum_{z^j_i\in\Zz^d\setminus\{0\}}\hat E_L(z^j_i)\right)\prod_{k=1}^N \delta_{Z_k,0}.
\eea
For $N=0,1,2$
\[
Q_{0,L}=1,\quad Q_{1,L}=L^d/\lambda_\beta^{d}, \quad Q_{2,L}=\frac{L^{2d}}{2\lambda_\beta^{2d}}\hat E_L(0).
\]
Because $Z_k=0$ for $k=1,\dots,N$ must be satisfied, not all the $N(N-1)/2$ variables $z^j_i$ can be chosen freely. Those taking zero do not influence the value of $Z_k$'s, so one must focus on the set of nonzero $z^j_i$. Given $\{z^j_i\in\Zz^d|1\leq i<j\leq N\}$, consider therefore a graph $G_{\{z^j_i\}}$ of $N$ vertices labeled from 1 to $N$ and an edge between $i$ and $j$ if $z^j_i\neq 0$.  $G_{\{z^j_i\}}$ divides into clusters (components, in graph theory). Vertex $k$ is a single-point cluster if $z^k_i=0$ and $z^j_k=0$ for all $i<k<j$. We call a nontrivial cluster valid if $Z_k=0$ can be solved for each $k$ in its vertex set with all the edge variables taking a nonzero value. A graph is valid if all of its nontrivial clusters are valid. Below we use the standard notation $V(G)$ and $E(G)$ for the  sets of vertices and edges, respectively, of a graph $G$, and $|V(G)|$ and $|E(G)|$ for the number of elements in these sets.

\vspace{5pt}
\noindent
{\em Lemma.} A nontrivial cluster is valid if and only if it is cyclic, i.e., every edge is at least in one cycle.
\vspace{5pt}

\noindent
{\em Proof.} The two-point cluster is not valid: $\{1, 2\}$ has the single edge $(12)$, and $Z_1=Z_2=0$ only if $z^2_1=0$. First we prove that
$n\geq 3$-point cycles are valid. Let $(l_1,\dots,l_n)$ be a cycle whose consecutive vertices are labeled by $n$ different positive integers $l_1,\dots,l_n$. Let $x_{i,i+1}$
denote the edge variable on the edge $(l_il_{i+1})$: $x_{i,i+1}=z^{l_{i+1}}_{l_i}$ if $l_{i+1}>l_i$ and $x_{i,i+1}=z^{l_{i}}_{l_{i+1}}$ if $l_{i+1}<l_i$. The equation $Z_{l_i}=0$ ($i=1,\dots,n$) is one of
\[
(1)\ - x_{i-1,i}-x_{i,i+1}=0\quad \mbox{if}\quad l_i<l_{i-1},l_{i+1},\quad
(2)\ x_{i-1,i}-x_{i,i+1}=0\quad \mbox{if}\quad l_{i-1}<l_i<l_{i+1},
\]
\[
(3)\ \ x_{i-1,i}+x_{i,i+1}=0\quad \mbox{if}\quad l_i>l_{i-1},l_{i+1},\quad
(4) -x_{i-1,i}+x_{i,i+1}=0\quad \mbox{if}\quad l_{i-1}>l_i>l_{i+1},
\]
where $x_{0,1}=x_{n,n+1}=x_{n,1}$. For clarity, we have written four equations instead of just two independent ones.
We see that whatever be the choice of, say, $x_{1,2}$, the other variables must take the same value with plus or minus sign. So there is at most a single free variable. To be definite, let $l_1=\min_j l_j$. Let us take an arbitrary nonzero vector $v$ and set $x_{1,2}=v$. $Z_{l_1}=0$ is equation (1) for $i=1$, it is solved with $x_{n,1}=-v$. We must check that going around the cycle there is no "frustration", all the equations can be solved. It is helpful to imagine an arrow on every edge, pointing towards the larger-labeled vertex. Vertex $i$ is called a source if $l_i<l_{i-1},l_{i+1}$ (two outgoing arrows), a sink if $l_i>l_{i-1},l_{i+1}$ (two incoming arrows), and a permeable point if $l_{i-1}<l_i<l_{i+1}$ or $l_{i-1}>l_i>l_{i+1}$ (one arrow in, one out).
In order to satisfy the respective equation, on different sides of a source (equation (1)) or a sink (equation (3)), $v$ must appear with different signs, while on both sides of a permeable point (equations (2) and (4)) the sign must be the same; see Fig. 1(a).
The problem is soluble because the number of sinks equals the number of sources. Specifically,
$i=1$ is a source; by passing it, $-v$ changes to $v$ while solving equation (1), stays $v$ by passing permeable points until the first sink and solves equation (2), passing the sink it changes to $-v$ while solving equation (3), stays $-v$ by passing permeable points and solves equation (4), and so on.

%%%    Figure begins  %%%%%%%%%%%%%%%%
\unitlength 1mm
\begin{picture}(35,35)(-20,-6)
\thicklines
\put(-15,11){\makebox(0,0)[l]{(a)}}
%   square
\put(0,0){\vector(0,1){12}} \put(0,12){\line(0,1){10}}
\put(0,0){\vector(1,0){12}} \put(12,0){\line(1,0){10}}
\put(0,22){\vector(1,0){12}} \put(12,22){\line(1,0){10}}
\put(22,0){\vector(0,1){12}} \put(22,12){\line(0,1){10}}
%    triangle
\put(32,0){\vector(0,1){12}} \put(32,12){\line(0,1){10}}
\put(32,0){\vector(2,1){12}} \put(44,6){\line(2,1){10}}
\put(54,11){\vector(-2,1){12}} \put(44,16){\line(-2,1){11.7}}
%               adding numbers to the vertices of the square
\put(-2,-3){\makebox(0,0)[lb]{1}}
\put(24,-3){\makebox(0,0)[rb]{2}}
\put(-2,25){\makebox(0,0)[lt]{4}}
\put(24,25){\makebox(0,0)[rt]{5}}
%               adding numbers to the vertices of the triangle
\put(30,-3){\makebox(0,0)[lb]{2}}
\put(56,11){\makebox(0,0){3}}
\put(30,25){\makebox(0,0)[lt]{5}}
%          adding +- v to the edges of the square
\put(11,-3){\makebox(0,0){v}}
\put(-3,11){\makebox(0,0)[b]{-v}}
\put(11,25){\makebox(0,0){-v}}
\put(19,11){\makebox(0,0){v}}
%         adding +-w to the edges of the triangle
\put(29,11){\makebox(0,0){-w}}
\put(46,4){\makebox(0,0)[r]{w}}
\put(46,18){\makebox(0,0)[r]{w}}
%%%%%%%%%%%%%%%%
%  square and triangle merged
%%%%%%%%%%%%%%%%
\put(80,11){\makebox(0,0)[r]{(b)}}
%   square
\put(91,0){\vector(0,1){12}} \put(91,12){\line(0,1){10}}
\put(91,0){\vector(1,0){12}} \put(103,0){\line(1,0){10}}
\put(91,22){\vector(1,0){12}} \put(103,22){\line(1,0){10}}
\put(113,0){\vector(0,1){12}} \put(113,12){\line(0,1){10}}
%   >
\put(113,0){\vector(2,1){12}} \put(125,6){\line(2,1){10}}
\put(135,11){\vector(-2,1){12}} \put(125,16){\line(-2,1){11.7}}
%   adding numbers to vertices
\put(89,-3){\makebox(0,0)[lb]{1}}
\put(114,-3){\makebox(0,0)[rb]{2}}
\put(89,25){\makebox(0,0)[lt]{4}}
\put(114,25){\makebox(0,0)[rt]{5}}
\put(137,11){\makebox(0,0){3}}
%   adding vectors to the edges
\put(102,-3){\makebox(0,0){v}}
\put(88,11){\makebox(0,0)[b]{-v}}
\put(102,25){\makebox(0,0){-v}}
\put(109,11){\makebox(0,0){v-w}}
\put(127,4){\makebox(0,0)[r]{w}}
\put(127,18){\makebox(0,0)[r]{w}}
\end{picture}
%%%%%   Figure ends   %%%%%%%%%%%%%%%

\hspace{35mm} {\em Figure 1.} Merging two cycles through a common edge.

\vspace{5mm}

For $m\geq 3$, let $(i_1\dots i_{m})$ be a cycle not equal to $(l_1,\dots,l_n)$ but
$\{i_1,\dots,i_{m}\}\cap\{l_1,\dots,l_n\}\neq\emptyset$, and take a vector $w\notin\{0,\pm v\}$. Following the procedure described above,
the equations $Z_{i_1}=\cdots=Z_{i_{m}}=0$ on $(i_1\dots i_{m})$ can be solved by assigning $y_{k,k+1}=\pm w$ in a suitable order to the edges $(i_ki_{k+1})$. Merge the cycles $(l_1\dots l_n)$ and $(i_1\dots i_{m})$ through their common vertices and edges. If two edges coincide, say, $(i_ki_{k+1})=(l_jl_{j+1})$, the new value of the edge variable is
\[
x_{j,j+1}+y_{k,k+1}\in \{v+w,v-w,-v+w,-v-w\};
\]
the other edges keep their individual value (Fig.1(b)). Clearly, $Z_k=0$ still holds for all $k\in \{i_1,\dots,i_{m}\}\cup\{l_1,\dots,l_n\}$, and all the edge variables are nonzero. The procedure can be continued by merging further cycles, and in a finite number of steps one can obtain any given connected finite cyclic graph.

In the opposite direction, suppose that $C$ is a valid cluster, and let a nonzero vector be assigned to each edge so that $Z_k=0$ for every $k\in V(C)$. Being valid, $C$ is without endpoints: if $k$ were an endpoint, it would have a single neighbor $j$, so $Z_k=z^k_j=0$ or $Z_k=z^j_k=0$ would be the only solution. Thus, $C$ contains cycles, and we can proceed by successive demerging. We choose a cycle $g$ in $C$ and prepare an image $g'$ of $g$ outside $C$, meaning that $g'$ has the same labeled vertices and the same edges as $g$. Next, we choose an edge $e$ of $g$ and, if it carries the vector $x$, we assign $-x$ to the image $e'$ of $e$. This uniquely determines the variables on the other edges of $g'$ in such a way that the constraints are satisfied. Now we merge $g'$ with $C$. As a result, the new vector on $e$ is zero, and $x$ or $-x$ must be added to the vectors on the other edges of $g$. By dropping all the edges from the merger whose new vector is zero and by dropping also the vertices that become isolated, we obtain a new graph, not necessarily connected, having at least one edge less than $C$, and still being valid. In a finite number of steps we can eliminate all the cycles of $C$. If the rest is nonempty, it cannot be valid, which is a contradiction. So $C$ is cyclic.   $\blacksquare$

\vspace{5pt}
The two products in Eq.~(\ref{partition-1}) split into factors according to the clusters. We rearrange the sum constituting $Q_{N,L}$ by summing first over the clusters containing vertex~1. Due to permutation invariance, for $n$-point clusters it suffices to consider the vertex set $\{1,2,\dots,n\}$ and to take into account all the possible partners of 1 in a cluster of $n$ points by a factor ${N-1\choose n-1}$. Let $\Sigma_n$ denote the family of valid $n$-point clusters of vertex set $\{1,2,\dots,n\}$. The only element of $\Sigma_1$ is $\{1\}$, $\Sigma_2$ is empty. For $n\geq 3$ the number of edges are between $n$ and $n(n-1)/2$. The clusters with $n$ edges are cycles, their number is $(n-1)!/2$, and there is a single cluster of $n(n-1)/2$ edges, the complete $n$-graph.
Separating in $Q_{N,L}$ the contribution of clusters of order $n$ containing~1, there remains $Q_{N-n,L}$. With $\rho=N/L^d$,
\begin{eqnarray}\label{QNL-recurrence}
Q_{N,L}&=&\frac{L^{dN}}{\lambda_\beta^{dN}N!}\sum_{n=1}^N{N-1\choose n-1}
\hat E(0)^{n(N-n)}
\sum_{C\in\Sigma_n}\hat E_L(0)^{{n\choose 2}-|E(C)|}
\sum_{\{z^j_i\in\Zz^d\setminus\{0\}|(ij)\in E(C)\}}\left(\prod_{(ij)\in E(C)}\hat E_L(z^j_i)\right)\prod_{i=1}^n\delta_{Z_i,0}
\nonumber\\
&\times&\sum_{\{z^j_i\in\Zz^d|n+1\leq i<j\leq N\}}\left(\prod_{n+1\leq i<j\leq N}\hat E_L(z^j_i)\right)
\prod_{k=n+1}^N \delta_{Z_k,0}
\nonumber\\
&=&\frac{1}{\rho\lambda_\beta^d}\sum_{n=1}^N \hat E(0)^{n(N-n)}
q_n^L Q_{N-n,L},
\end{eqnarray}
where $q^L_1=1$, $q^L_2=0$, and for $n\geq 3$,
\be
q^{L}_n=\frac{L^{d(n-1)}}{\lambda_\beta^{d(n-1)}(n-1)!}
\sum_{C\in\Sigma_n}\hat E_L(0)^{{n\choose 2}-|E(C)|}
\sum_{\{z^j_i\in\Zz^d\setminus\{0\}|(ij)\in E(C)\}}\left(\prod_{(ij)\in E(C)}\hat E_L(z^j_i)\right)\prod_{i=1}^n\delta_{Z_i,0}.
\ee
The factor $\hat E(0)^{n(N-n)}$ comes from $z^k_j=0$ when $j\leq n<k\leq N$.
Observe that $q^L_n$ is independent of $N$ and can be negative.
Equation~(\ref{QNL-recurrence}) is the recurrence relation announced in the title of the section.

\section{Thermodynamic limit}

By rearranging Eq.~(\ref{QNL-recurrence}),
\begin{eqnarray*}
\rho\lambda_\beta^{d}
=\sum_{n=1}^N \hat E_L(0)^{n(N-n)}
q^{L}_n\frac{Q_{N-n,L}}{Q_{N,L}}.
\end{eqnarray*}
The equality must hold when $N, L\to\infty$ while $N/L^d=\rho$ is fixed.
For a positive integer $M<N$ we cut the sum into two parts, for $n\leq M$ and $M+1\leq n\leq N$.
Taking the thermodynamic limit,
\be\label{full}
\rho\lambda_\beta^{d}=\sum_{n=1}^M \lim_{N,L\to\infty, N/L^d=\rho}\hat E_L(0)^{n(N-n)}
q^{L}_n\frac{Q_{N-n,L}}{Q_{N,L}}
+\lim_{N,L\to\infty, N/L^d=\rho}\sum_{n=M+1}^N \hat E_L(0)^{n(N-n)}
 q^{L}_n\frac{Q_{N-n,L}}{Q_{N,L}}.
\ee
The limit could be interchanged with the summation up to $M$ and, as we will see, it exists for fixed $n$. It follows that the $N,L\to\infty$ limit of the sum from $M+1$ to $N$ also exists. Suppose now that by sending $M$ to infinity we find
\be\label{no-infinite-cluster}
\lim_{M\to\infty}\lim_{N,L\to\infty, N/L^d=\rho}\sum_{n=M}^N \hat E_L(0)^{n(N-n)}
 q^{L}_n\frac{Q_{N-n,L}}{Q_{N,L}}=0.
\ee
Were it not zero, it would be due to infinite clusters. Equation (\ref{no-infinite-cluster}) implies
\be\label{finite-contribution}
\rho\lambda_\beta^{d}=\sum_{n=1}^\infty\lim_{N,L\to\infty, N/L^d=\rho} \hat E_L(0)^{n(N-n)}
q^{L}_n\frac{Q_{N-n,L}}{Q_{N,L}},
\ee
finite clusters carry the whole density. The task is to show that Eqs.~(\ref{no-infinite-cluster}) and (\ref{finite-contribution}) are true if $\rho\lambda_\beta^{d}$ is small, but fail if $\rho\lambda_\beta^{d}$ is above a critical value, so infinite clusters contribute to $\rho\lambda_\beta^{d}$.

Defining $f_{N,L}(\beta)$ via $Q_{N,L}=\exp\{-\beta L^d f_{N,L}(\beta)\}$,
\[
\lim_{N,L\to\infty, N/L^d=\rho}f_{N,L}(\beta)=f(\rho,\beta)
\]
exists~\cite{R2}. We have
\be\label{fugacity}
\frac{{Q}_{N-1,L}}{{Q}_{N,L}}
=\exp\left\{\beta\frac{f_{N,L}(\beta)-f_{N-1,L}(\beta)}{N/L^d-(N-1)/L^d}\right\}\to \exp\left\{\beta\frac{\partial {f}(\rho,\beta)}{\partial \rho}\right\},
\ee
 therefore the thermodynamic limit of $Q_{N-n,L}/Q_{N,L}$ is
\be\label{Q_{N-n}/Q_N}
\lim_{N,L\to\infty, N/L^d=\rho} \frac{{Q}_{N-n,L}}{{Q}_{N,L}}
=\!\!\lim_{N,L\to\infty, N/L^d=\rho}\prod_{i=1}^n \frac{{Q}_{N-i,L}}{{Q}_{N-i+1,L}}
=\exp\left\{n\beta\frac{\partial {f}(\rho,\beta)}{\partial \rho}\right\}.
\ee
$f(\rho,\beta)$ is the free energy density, $\partial f(\rho,\beta)/\partial\rho$ is the chemical potential, and $\exp\{\beta\partial f(\rho,\beta)/\partial\rho\}$ is the activity.
Furthermore,
\[
\hat E_L(0)^{n(N-n)}=\left[\frac{1}{L^d}\int_\Lambda E_L(x)\d x \right]^{n(N-n)}
=\left[1-\frac{1}{L^d}\int_\Lambda v_L(x)\d x\right]^{n(N-n)},
\]
and therefore
\be\label{E0-to-n(N-n)}
\lim_{N,L\to\infty, N/L^d=\rho} \hat E_L(0)^{n(N-n)}=e^{-n\rho\hat v(0)}.
\ee
The chemical potential is determined by $\rho\lambda_\beta^{d}$ and the sequence
\[
q_n=\lim_{L\to\infty}q^{L}_n
\]
as the solution for $\mu$ of the equation
\be\label{first-eq-for-mu}
A_{\rho,\beta}(\mu):=\sum_{n=1}^\infty q_n e^{n(\beta\mu-\rho\hat v(0))}
=\rho\lambda_\beta^{d}.
\ee
For nonnegative pair potentials equations~(\ref{partition-function}) and (\ref{fugacity}) yield the trivial bounds
\[
\rho\lambda_\beta^{d}\leq e^{\beta\partial f(\rho,\beta)/\partial\rho}\leq \rho\lambda_\beta^{d}\,e^{\beta\rho\int u\d x}.
\]
The lower bound requires $u\geq 0$, the upper bound is always true due to Jensen's inequality. Note that $\int u\d x\geq 0$ holds for stable interactions, even if $u$ is partly negative.

Let us see, how infinite clusters can emerge at high density. $A_{\rho,\beta}(\mu)$ must be finite for the equation~(\ref{first-eq-for-mu}) to be solved. The sum is absolutely convergent if $\mu< \bar{\mu}(\rho,\beta)$, where $e^{\beta \bar{\mu}(\rho,\beta)}$ is the convergence radius of the power series $\sum_n e^{-n\rho \hat v(0)} q_n z^n$. Stability of the interaction implies that $|q_n|$ cannot increase faster than exponentially with $n$, therefore the convergence radius is nonzero, $\bar{\mu}(\rho,\beta)>-\infty$. Fixing any $\beta<\infty$, for $\rho$ sufficiently small,  Eq.~(\ref{first-eq-for-mu}) can be solved for $\mu$. If $\mu(\rho,\beta)$ denotes the solution, then $A_{\rho,\beta}(\mu)=0$ yields $\mu(0,\beta)=-\infty$ and, due to the convexity of $f(\rho,\beta)$ as a function of $\rho$~\cite{R2}, $\mu(\rho,\beta)=\partial f(\rho,\beta)/\partial\rho$ strictly increases with $\rho$ in an interval $(0,\rho_c(\beta))$, where $\rho_c(\beta)\leq\infty$.
Taking $\mu$ as an independent variable, $A_{\rho,\beta}(\mu)$ strictly increases with it in the interval $(-\infty,\bar{\mu}(\rho,\beta))$, and tends to $A_{\rho,\beta}(\bar{\mu}(\rho,\beta))\leq \infty$. If $A_{\rho,\beta}(\bar{\mu}(\rho,\beta))=\infty$ then $\rho_c(\beta)=\infty$. If $A_{\rho,\beta}(\bar{\mu}(\rho,\beta))$ is finite, $\rho_c(\beta)$ is also finite. Infinite clusters only appear in this case, for $\rho>\rho_c(\beta)$, signaling that finite clusters cannot carry a density greater than $\rho_c(\beta)$.

Now whether $A_{\rho,\beta}(\bar{\mu}(\rho,\beta))$ is finite or not depends on the asymptotic form of $q_n$. If
\be\label{epsilon}
\epsilon(\beta)=-\beta^{-1}\lim_{n\to\infty}n^{-1}\ln|q_n|<\infty,
\ee
for $n\geq 3$ we can write
\[
q_n=\alpha_n e^{-n\beta[\epsilon(\beta)+\delta_n(\beta)]},
\]
where $\alpha_n\in\{\pm 1\}$, $\delta_n(\beta)\to 0$ if $n\to\infty$, and $\epsilon(\beta)$ and $\delta_n(\beta)$ can be positive or negative. With the above form of $q_n$,
\be\label{Arhobeta_detailed}
A_{\rho,\beta}(\mu)=e^{\beta\mu-\rho\hat v(0)}+\sum_{n=3}^\infty\alpha_n e^{-n\beta\delta_n(\beta)} e^{n[\beta\mu-\beta\epsilon(\beta)-\rho\hat v(0)]}\qquad (q_1=1, q_2=0).
\ee
Because $n\delta_n(\beta)=o(n)$, $\beta\mu-\beta\epsilon(\beta)-\rho\hat v(0)=0$ determines the convergence radius via
\be\label{barmu}
\bar{\mu}(\rho,\beta)=\epsilon(\beta)+\rho\hat v(0)/\beta,
\ee
so
\be\label{A_beta(barmu-beta)}
\bar A_\beta:=A_{\rho,\beta}(\bar{\mu}(\rho,\beta))=e^{\beta\epsilon(\beta)}+\sum_{n=3}^\infty\alpha_n e^{-n\beta\delta_n(\beta)},
\ee
independent of $\rho$!
We see that the fate of infinite clusters depends on the correction $e^{o(n)}$ to the exponential decay or growth of $|q_n|$; the whole difficulty of proving the transition lies in this fact. In this respect, the appearance of infinite clusters is analogous to the appearance of infinite permutation cycles in systems of interacting bosons~\cite{S2}.

It cannot be ruled out that for certain interactions
$
|q_n|\sim \exp\{-c_1 n^{1+c_2}\}
$
where $c_1,\, c_2>0$. If this holds true, the convergence radius of the series $A_{\rho,\beta}(\mu)$ is infinite, there are only finite clusters, and there are no classical phase transitions at all densities. This may be the case for long-range repulsive forces which, as previously noted~\cite{S2}, can also prevent the occurrence of infinite permutation cycles and thus of Bose-Einstein condensation.

One general characteristic remains to be mentioned. Phase diagrams, whether experimental or theoretical, exhibit a critical point $(\rho^*=\rho_c(\beta^*),\beta^*)$
on the $(\rho,\beta)$ plane, such that for $\beta<\beta^*$ the gas does not condense and only a single fluid phase exists. In our study this translates into the fact that the function $\bar A_\beta$ tends towards infinity as $\beta$ approaches a certain value $\beta^*>0$ from above. For $\beta<\beta^*$, Eq.~(\ref{first-eq-for-mu}) can be solved with $\mu<\bar{\mu}(\rho,\beta)$, where $\mu-\bar{\mu}(\rho,\beta)<0$ approaches zero as $\rho$ increases.
% with gases, van der Waals theory, and numerical work on Lennard-Jones systems~\cite{SK},~\cite{SSH} tell us that

If $\bar A_\beta$ is finite, the critical density is given by
\[
\rho_c(\beta)=\bar A_\beta/\lambda_\beta^d.
\]
The above argument applies to interactions without a hard core. Now $u$ with a hard core is allowed if we apply a mollifier to the Boltzmann factor $e^{-\beta u(x)}$. In this case $\rho_c(\beta)\leq\rho_{\rm cp}$, the close-packing density, and the condition for the appearance of infinite clusters is $\bar A_\beta/\lambda_\beta^d<\rho_{\rm cp}$. $A_{\rho,\beta}(\mu)$ cannot be continued analytically beyond $\rho_{\rm cp}\lambda_\beta^d$, so the alternative is $\bar A_\beta/\lambda_\beta^d=\rho_{\rm cp}$, no infinite cluster for $\rho<\rho_{\rm cp}$.

Note that for $\rho<\rho_c(\beta)$, $e^{-n\rho\hat v(0)}q_n e^{n\beta \partial f/\partial\rho}$ decays at least exponentially fast,
%$\partial A_{\rho,\beta}(\mu)/\partial\mu\neq 0$,
$f(\rho,\beta)$ is a real analytic function of $\rho$,
and from Eq.~(\ref{first-eq-for-mu}) and the implicit function theorem a relation between the first and second $\rho$-derivative of $f(\rho,\beta)$ can be derived. Let
\[
F(\rho,\mu):=\rho\lambda_\beta^d-A_{\rho,\beta}(\mu),
\]
Differentiating $F(\rho,\mu(\rho))=0$ with respect to $\rho$ gives
\[
\frac{d\mu}{d\rho}=\frac{\lambda_\beta^d-\partial A_{\rho,\beta}(\mu)/\partial\rho}{\partial A_{\rho,\beta}(\mu)/\partial\mu}.
\]
Substituting from Eq.~(\ref{first-eq-for-mu}):
\[
\partial A_{\rho,\beta}(\mu)/\partial\rho=-\hat v(0)\sum_{n=1}^\infty n q_n e^{-n\rho\hat v(0)} e^{n\beta\mu},\quad
\partial A_{\rho,\beta}(\mu)/\partial\mu=\beta\sum_{n=1}^\infty n q_n e^{-n\rho\hat v(0)} e^{n\beta\mu},\quad \mu=\partial f(\rho,\beta)/\partial\rho,
\]
we obtain
\begin{eqnarray}\label{2nd-derivative -of-f}
\frac{\partial^2 f(\rho,\beta)}{\partial\rho^2}
=\frac{\lambda_\beta^d}{\beta\sum_{n=1}^\infty n q_n e^{-n\rho\hat v(0)} e^{n\beta\partial f(\rho,\beta)/\partial\rho}}+\frac{\hat v(0)}{\beta}.
\end{eqnarray}
The interest in $\partial^2 f(\rho,\beta)/\partial\rho^2$ lies in its relation with the $\rho$-derivative of the pressure:
\be\label{rho-derivative -of-p}
p(\rho,\beta)=\rho\, \frac{\partial f(\rho,\beta)}{\partial\rho} – f(\rho,\beta),\quad \frac{\partial p(\rho,\beta)}{\partial\rho} = \rho \frac{\partial^2 f(\rho,\beta)}{\partial\rho^2}.
\ee
For densities above $\rho_c(\beta)$, $\partial f(\rho,\beta)/\partial\rho$ remains a monotonically increasing, although not necessarily strictly increasing, function of $\rho$, cf. Section~\ref{condensed}.

We now derive the $L\to\infty$ limit of $q^L_n$. As noted above, $q_1=1$ and $q_2=0$. For $n\geq 3$ we proceed as follows.
Substituting $E_L(x)=1-v_L(x)$ in the formula for $\hat E_L(z)$,
\[
\hat E_L(z)=\delta_{z,0}-\frac{1}{L^d}\int_\Lambda e^{-\i \frac{2\pi}{L}z\cdot x}v_L(x)\d x,
\]
from which we obtain, due to the integrability of $v$,
\[
\lim_{L\to\infty}\hat E_L(0)^{{n\choose 2}-|E(C)|}=\lim_{L\to\infty}\left[1-\frac{1}{L^d}\int_\Lambda v_L(x)\d x\right]^{{n\choose 2}-|E(C)|}=1.
\]
Note the contrast with Eq.~(\ref{E0-to-n(N-n)}).
Thus,
\[
q_n=\frac{1}{\lambda_\beta^{d(n-1)}(n-1)!}\sum_{C\in\Sigma_n}\lim_{L\to\infty}L^{d(n-1)}
\sum_{\{z^j_i\in\Zz^d\setminus\{0\}|(ij)\in E(C)\}}\left(\prod_{(ij)\in E(C)}\hat E_L(z^j_i)\right)\prod_{i=1}^n\delta_{Z_i,0}.
\]
Here
\[
\prod_{(ij)\in E(C)}\hat E_L(z^j_i)
=(-1)^{|E(C)|}(L^d)^{-|E(C)|}\prod_{(ij)\in E(C)}\int_\Lambda e^{-\i \frac{2\pi}{L}z^j_i\cdot x}v_L(x)\d x.
\]
The limit $L\to\infty$ is taken in two steps. First, the integrals can be replaced with
$
\hat v(z^j_i/L),
$
keeping the $L$-dependence in the exponent, allowed because
\[
\int_\Lambda e^{-\i 2\pi\kappa \cdot x}v_L(x)\d x\stackrel{L\to\infty}{\longrightarrow} \hat v(\kappa)
\]
uniformly for $\kappa$ in any bounded interval. This yields
\[
q_n=\frac{1}{(n-1)!}\sum_{C\in\Sigma_n}\frac{(-1)^{|E(C)|}}{\lambda_\beta^{d(n-1)}}
\lim_{L\to\infty}(L^d)^{n-1-|E(C)|}
\sum_{\{z^j_i\in\Zz^d\setminus\{0\}|(ij)\in E(C)\}}\left(\prod_{(ij)\in E(C)}\hat v(z^j_i/L)\right)\prod_{i=1}^n\delta_{Z_i,0}.
\]
The dependence on $L^d$ shows that among $\{z^j_i\in\Zz^d\setminus\{0\}|(ij)\in E(C)\}$ there are $N_{\rm free}(C)=|E(C)|-n+1$ free summation variables, and the residual $n-1$ variables are linear combinations of them. In the limit of $L$ going to infinity, $L^{-d(|E(C)|-n+1)}$ times the multiple sum over $\{z^j_i\in\Zz^d\setminus\{0\}|(ij)\in E(C)\}$ becomes a multiple integral with respect to the free variables,
\be\label{q_n}
q_n
=\frac{1}{(n-1)!}\sum_{C\in\Sigma_n}\frac{(-1)^{|E(C)|}}{\lambda_\beta^{d(n-1)}}
\int\prod_{i=1}^{|E(C)|}\hat v\left(\sum_{j=1}^{|E(C)|-n+1}c_{ij}\kappa_j\right)\d\kappa_1\dots\d\kappa_{|E(C)|-n+1}
=\frac{1}{(n-1)!}\sum_{C\in\Sigma_n} g(C),
\ee
where $c_{ij}\in\{0,\pm 1\}$, and for any $i$, $c_{ij}\neq 0$ for at least one $j$. Here $i$ labels an edge and $j$ may be assigned to an edge or a cycle, see below. The transition from sums to integrals relies on $\hat{v}\in C\cap L^1(\Rr^d)$.

Recall that a block of a graph is a maximal connected subgraph that remains connected if any single vertex is removed. For $n\geq 3$ an element of $\Sigma_n$ is a block or a tree of blocks with a total of $n$ labeled vertices, each block being a cycle or a merger of cycles through edges. Because the blocks do not share edges, the variables associated with different blocks are independent.
Therefore, if $C=(B_1,\dots,B_l)$ denotes any tree of blocks $B_1,\dots,B_l$, then $N_{\rm free}(C)=\sum_{k=1}^l N_{\rm free}(B_k)$, that is,
\[
|E(B_1,\dots,B_l)|-|V(B_1,\dots,B_l)|+1=\sum_{k=1}^l \left(|E(B_k)|-|V(B_k)|+1\right).
\]
This is in accordance with the blocks forming a tree,
\be\label{V(oB)}
 |E(B_1,\dots,B_l)|=\sum_{k=1}^{l}|E(B_k)| \quad\mbox{and}\quad |V(B_1,\dots, B_l)|=1+\sum_{k=1}^l \left( |V(B_k)|-1\right).
\ee
The summand in (\ref{q_n}) divides into a product over the blocks. With  (\ref{V(oB)}),
$
g(C)=\prod_{k=1}^l g(B_k)
$
where
\[
g(B)=
\frac{(-1)^{|E(B)|}}{\lambda_\beta^{d(|V(B)|-1)}}\int\prod_{i=1}^{|E(B)|}\hat v\left(\sum_{j=1}^{|E(B)|-|V(B)|+1}c_{ij}\kappa_j\right)\d\kappa_1\dots\d\kappa_{|E(B)|-|V(B)|+1}.
\]
The dimensionless $g(B)$ is expressed in terms of dimensional quantities, the length $\lambda_\beta$, the inverse lengths $\kappa_{j1},\dots,\kappa_{jd}$ (components of $\kappa_j$), the volume $\hat v$. This can be changed by introducing
\be\label{vbeta}
v_\beta(y)=v(\lambda_\beta y)=1-e^{-\beta u(\lambda_\beta y)},
\ee
where $y\in\Rr^d$ is now dimensionless. With the variable change $\kappa'=\lambda_\beta\kappa$,
\[
\frac{\hat v(\kappa)}{\lambda_\beta^{d}}=\int_{\Rr^d} e^{-\i 2\pi\kappa'\cdot y}v_\beta(y)\d y=\widehat{v_\beta}(\kappa')
\]
and, by dropping the prime,
\bea\label{g(B)}
g(B)=
(-1)^{|E(B)|}\int\prod_{i=1}^{|E(B)|}\widehat{v_\beta}\left(\sum_{j=1}^{|E(B)|-|V(B)|+1}c_{ij}\kappa_j\right)\d\kappa_1\dots\d\kappa_{|E(B)|-|V(B)|+1}.
\eea
The new $\kappa_j$ and $\widehat{v_\beta}$ are dimensionless. What we gained is that the entire $\beta$-dependence is now in $\widehat{v_\beta}$. Due to temperedness, $\beta|u(\lambda_\beta y)|<c\beta/(\lambda_\beta |y|)^\eta$ for $\lambda_\beta|y|>R$.
In $d\geq 2$ dimensions $\eta>d$ implies that for any $y\neq 0$, $v_\beta(y)$ goes to zero with $\beta$ going to infinity. From $v_\beta\in L^1(\Rr^d)$ and $|\widehat{v_\beta}(\kappa)|\leq\int|v_\beta(y)|\d y$ we conclude that $g(B)$ and, hence, $q_n$ for $n\geq 3$ also go to zero.
To summarize, $q_1=1$, $q_2=0$ and
\be
q_n=\frac{1}{(n-1)!}\sum_{C\in\Sigma_n}\prod_{B\subset C} g(B), \qquad n\geq 3.
\ee
\vspace{5pt}
\noindent
{\em Remark 1.} If a block is a planar graph or a graph consisting of the vertices and edges of a convex polyhedron, $N_{\rm free}$ is related to the Euler characteristic $V+F-E=2$. Comparison with it shows that $N_{\rm free}=F-1$, where $F$ is the number of faces (the exterior face included for planar graphs). We choose any $F-1$ faces of the polyhedron, or all the interior faces of the planar block, and assign a free variable to each bordering cycle. Referring to Eq.~(\ref{q_n}), $c_{ij}\neq 0$ for a single $j$ if $i$ is an external edge and for two $j$'s if $i$ is an internal edge.

\vspace{5pt}
\noindent
{\em Remark 2.} For the complete $n$-graph the general solution of the system of equations $Z_k=0$, $k=1,\dots,n$, is easily obtained. In this case $N_{\rm free}=n(n-1)/2-n+1=(n-1)(n-2)/2$, the number of edges of the complete $(n-1)$-graph. The variables of this latter, $z^k_j$ for $1\leq j<k\leq n-1$ can be taken to be free; then, with
\[
z^n_l=\sum_{j=1}^{l-1}z^l_j-\sum_{k=l+1}^{n-1}z^k_l  \qquad l=1,\dots,n-1,
\]
all the equations will be satisfied. The integration variables are $\kappa^k_j$, $1\leq j<k\leq n-1$.

\vspace{5pt}
\noindent
{\em Remark 3.} For an arbitrary block $B$ finding $N_{\rm free}(B)=|E(B)|-|V(B)|+1$ edges that can carry free variables may be difficult. It is easier to find $N_{\rm free}(B)$ cycles that cover all the edges of $B$, and assign a free variable to each cycle. The variable is then equipped with edge-dependent signs along the cycle. After merging the cycles, on every edge we obtain the sum of the signed variables belonging to the cycles incident on that edge. The choice of the generating cycles is not unique, but the passage between two sets involves a variable change whose Jacobian is of modulus one, so the result is unique.

\vspace{5pt}
If $B$ is the complete $n$-graph, the integral in $g(B)$ is
\[
\int\left[\prod_{1\leq j<k \leq n-1}\widehat{v_\beta}(\kappa^k_j)\right]\cdot \left[\prod_{l=1}^{n-1}\widehat{v_\beta}\left(\sum_{j=1}^{l-1}\kappa^l_j-\sum_{k=l+1}^{n-1}\kappa^k_l\right)\right]\prod_{1\leq j<k \leq n-1}\d\kappa^k_j.
\]
If $B$ is a $n$-cycle, the integral is
$
\int \widehat{v_\beta}(\kappa)^n\d\kappa
$
(because $\widehat{v_\beta}(\kappa)=\widehat{v_\beta}(-\kappa)$). Due to $v_\beta, \widehat{v_\beta}\in L^1(\Rr^d)$,
\bea\label{singlecycle}
\int \widehat{v_\beta}(\kappa)^n\d\kappa
&=&\int \widehat{(v_\beta\ast\cdots\ast v_\beta)} (\kappa)\d\kappa=(v_\beta\ast\cdots\ast v_\beta)(0)
\nonumber\\
&=&\int v_\beta(x_1)\dots v_\beta(x_{n-1})v_\beta(-x_1-\cdots-x_{n-1})\d x_1\dots\d x_{n-1}
\eea
and the integral is positive if $u\geq 0$ (because then $v_\beta\geq 0$) or $\widehat{v_\beta}\geq 0$. In such a case the sign of $g(B)$ is $(-1)^{|E(B)|}$.

\section{An illustrative example}\label{example}

We choose a pair potential defined by
$
\hat v(\kappa)=\lambda^d e^{-\pi\lambda^2\kappa^2};
$
then $v(x)=e^{-\pi x^2/\lambda^2}$ and
\[
\beta u(x)=-\ln\left(1-e^{-\pi x^2/\lambda^2}\right).
\]
This is purely repulsive, its asymptotic forms are
\[
\beta u(x)\sim\left\{\begin{array}{rl}
e^{-\pi x^2/\lambda^2}, & x\to \infty\\
-\ln\pi x^2/\lambda^2, & x\to 0.
\end{array}\right.
\]
$u$ depends on $\beta$, but as long as we only look at the $\rho$-dependence, the interaction is legitimate.
Let us retain from $\Sigma_n$ only the $(n-1)!/2$ cycles of length $n$. There are many more elements of $\Sigma_n$, and one can only hope that, with the cycles only, the result is qualitatively correct. There is no phase transition in one dimension for short-range repulsive interactions, so we focus on $d\geq 2$.
For $n\geq 3$
\[
\int \hat v(\kappa)^n\d\kappa=\frac{\lambda^{d(n-1)}}{n^{d/2}}
\]
and
\[
q_n\approx \frac{(-1)^n}{2n^{d/2}}(\lambda/\lambda_\beta)^{d(n-1)}.
\]
With the approximative $q_n$,
\bea
A_{\rho,\beta}(\mu)=\sum_{n=1}^\infty e^{-n\rho\hat v(0)}q_n e^{n\beta\mu}
= e^{\beta\mu-\rho\hat v(0)}+\frac{1}{2}\left(\frac{\lambda_\beta}{\lambda}\right)^d\sum_{n=3}^\infty \frac{(-1)^n}{n^{d/2}}e^{n[\beta\mu-\rho\hat v(0)-d\ln(\lambda_\beta/\lambda)]}
\nonumber\\
=e^{\beta\mu-\rho\hat v(0)}\left[1+\frac{1}{2}\sum_{n=3}^\infty \frac{(-1)^n}{n^{d/2}}e^{(n-1)[\beta\mu-\rho\hat v(0)-d\ln(\lambda_\beta/\lambda)]}\right].
\eea
If $\beta\mu< d\ln(\lambda_\beta/\lambda)+\rho\hat v(0)$, the infinite sum is absolutely convergent.
If $\beta\mu= d\ln(\lambda_\beta/\lambda)+\rho\hat v(0)$, it is still absolutely convergent for $d\geq 3$, nonabsolutely convergent for $d=1,2$, and for any $d\geq 1$ it
is bounded below by $-3^{-d/2}$, so
\[
0<e^{\beta\mu-\rho\hat v(0)}\left[1-\frac{1}{2\cdot 3^{d/2}}\right]< A_{\rho,\beta}(\mu)<\infty,\quad \beta\mu\leq d\ln(\lambda_\beta/\lambda)+\rho\hat v(0).
\]
If $\beta\mu>d\ln(\lambda_\beta/\lambda)+\rho\hat v(0)$, $A_{\rho,\beta}(\mu)$ is divergent, therefore $\bar\mu(\rho,\beta)=[d\ln(\lambda_\beta/\lambda)+\rho\hat v(0)]/\beta$. Because
\[
\bar A_\beta=\left(\frac{\lambda_\beta}{\lambda}\right)^d\left[1+\frac{1}{2}\sum_{n=3}^\infty \frac{(-1)^n}{n^{d/2}}\right]
\]
is convergent, by Abel's theorem $A_{\rho,\beta}(\mu)$ tends to it as $\mu$ tends to $\bar\mu(\rho,\beta)$.
It remains for us to verify the consistency of the approximation. First, $A_{\rho,\beta}(\mu)$ must monotonically increase with $\mu$ for $-\infty<\mu<\bar{\mu}(\rho,\beta)$; second, $\lim_{\mu\uparrow\bar\mu(\rho,\beta)}\d A_{\rho,\beta}(\mu)/\d\mu$ must exist, so that $\partial p(\rho,\beta)/\partial\rho$ has a left limit at $\rho_c(\beta)$, cf. Eqs.~(\ref{2nd-derivative -of-f}), (\ref{rho-derivative -of-p}). By introducing $\gamma=-\beta[\mu-\bar\mu(\rho,\beta)]$, $A_{\rho,\beta}(\mu)=(\lambda_\beta/\lambda)^d B(\gamma)$, where
\[
B(\gamma)=e^{-\gamma}+\frac{1}{2}\sum_{n=3}^\infty \frac{(-1)^n}{n^{d/2}}e^{-n\gamma}.
\]
For $\gamma>0$, $B(\gamma)$ must be a strictly decreasing function, and $\lim_{\gamma\downarrow 0}\d B(\gamma)/\d\gamma$ must exist. Now
\[
\frac{\d B(\gamma)}{\d\gamma}=-\left[e^{-\gamma}+\frac{1}{2}\sum_{n=3}^\infty \frac{(-1)^n}{n^{d/2-1}}e^{-n\gamma}\right]
=-\left[e^{-\gamma}-\frac{e^{-3\gamma}}{2\cdot 3^{d/2-1}}\right]
-\sum_{4\leq n\,{\rm even}}\left[\frac{e^{-n\gamma}}{n^{d/2-1}}-\frac{e^{-(n+1)\gamma}}{(n+1)^{d/2-1}}\right].
\]
We see that $d\geq 2$ dimensions pass both tests, $\d B/\d\gamma<0$ and has a finite negative limit at zero. So
\[
%A_{\rho,\beta}(\bar\mu(\beta))=\left(\frac{\lambda_\beta}{\lambda}\right)^d\left[1+\frac{1}{2}\sum_{n=3}^\infty \frac{(-1)^n}{n^{d/2}}\right],\quad
\rho_c
=\frac{\bar A_\beta}{\lambda_\beta^d}=\frac{1}{\lambda^d}B(0)
=\frac{1}{\lambda^d}\left[1+\frac{1}{2}\sum_{n=3}^\infty \frac{(-1)^n}{n^{d/2}}\right].
\]
If $\rho\leq \rho_c$, the equation $A_{\rho,\beta}(\mu)=\rho\lambda_\beta^d$ can be solved for $\mu$, and all the particles are in finite clusters. If $\rho>\rho_c$, there are infinite clusters, and the particle density in them is $\rho-\rho_c$.
The analytic form of the chemical potential changes at the transition point. Due to the repulsive interaction, there is no liquid state, for $d\geq 3$ the singularity can indicate the beginning of gas-solid coexistence. In two dimensions it may be related to the appearance of the hexatic phase -- or even be an artefact.

\section{No condensed phase without infinite clusters}\label{condensed}

Not knowing $q_n$ any better, we can nevertheless affirm that any condensed phase
must contain infinite clusters. Consider, for example, the vapor-liquid transition. Fix a subcritical temperature, so $\beta>\beta^*$, $\bar A_\beta<\infty$.
Increase the density of the vapor by increasing the pressure. At some $\rho=\rho_{v+}(\beta)$ condensation begins and, as we know from experiment, the van der Waals theory, and numerical simulations with, for example, the Lennard-Jones potential~\cite{SK, SSH}, the vapor of density $\rho_{v+}(\beta)$ and the liquid of density $\rho_{l-}(\beta)$ coexist for the global density in the interval $(\rho_{v+}(\beta),\rho_{l-}(\beta))$. The chemical potential is the same in both phases and is therefore constant throughout the coexistence interval. It is
$\mu(\rho_{v+}(\beta),\beta)=\epsilon(\beta)+\rho_{v+}(\beta)\hat v(0)/\beta$, cf. Eq.~(\ref{barmu}), where
\[
\rho_{v+}(\beta)=\bar A_\beta/\lambda_\beta^d.
\]
For $\rho>\rho_{v+}(\beta)$ we must look at the full equation (\ref{full}) which, by taking the limit of $M$ going to infinity, is
\be
\rho\lambda_\beta^d=A_{\rho,\beta}(\mu)+\lim_{M\to\infty}\lim_{N,L\to\infty, N/L^d=\rho}\sum_{n=M}^N \hat E_L(0)^{n(N-n)} q^{L}_n\frac{Q_{N-n,L}}{Q_{N,L}}.
\ee
$A_{\rho,\beta}(\mu)$ cannot be larger than $\bar A_\beta$, so the equation
\be\label{full-limit}
\lim_{M\to\infty}\lim_{N,L\to\infty, N/L^d=\rho}\sum_{n=M}^N\hat E_L(0)^{n(N-n)} q^{L}_n\frac{Q_{N-n,L}}{Q_{N,L}}
=(\rho-\rho_{v+}(\beta))\lambda_\beta^d
\ee
is valid for any $\rho\geq \rho_{v+}(\beta)$. Beyond some $\rho_{l-}(\beta)$ the chemical potential continues to increase with $\rho$, and the equation~(\ref{full-limit}) defines a condition on it. By further increasing the pressure, we arrive at the coexistence of the liquid of density $\rho_{l+}(\beta)$ and a solid of density $\rho_{s-}(\beta)$. In the coexistence interval $(\rho_{l+}(\beta),\rho_{s-}(\beta))$ the chemical potential is again constant at the value $\partial f(\rho,\beta)/\partial\rho|_{\rho=\rho_{l+}(\beta)}$, and continues to increase beyond $\rho=\rho_{s-}(\beta)$. The infinite clusters in a solid must be different from those in the liquid. In a crystal the atomic positions are correlated on long distances,
in the liquid the correlation is of medium range. This suggests that in a crystal the infinite clusters are macroscopic, while in the liquid they are submacroscopic. Accordingly, for $\rho\geq \rho_{v+}(\beta)$
\bea
\rho=\rho_{v+}(\beta)
+\lambda_\beta^{-d}
\lim_{M\to\infty}\lim_{N,L\to\infty, N/L^d=\rho}\sum_{n=M}^{\lfloor N/M\rfloor-1}\hat E_L(0)^{n(N-n)} q^{L}_n\frac{Q_{N-n,L}}{Q_{N,L}}
\nonumber\\
+\lambda_\beta^{-d}\lim_{M\to\infty}\lim_{N,L\to\infty, N/L^d=\rho}\sum_{n=\lfloor N/M\rfloor}^{N}\hat E_L(0)^{n(N-n)} q^{L}_n\frac{Q_{N-n,L}}{Q_{N,L}},
\eea
densities carried by finite, infinite submacroscopic, and macroscopic clusters. Submacroscopic clusters must be infinite in number to contribute to the density.  If $\beta<\infty$ then $\rho_{v+}(\beta)>0$, finite clusters appear along with infinite clusters, in liquids and crystals, at positive temperatures. On the contrary, $\lim_{\beta\to\infty}\rho_{v+}(\beta)$ can be positive or zero, and both results have an interesting consequence on the ground state, the $\beta\to\infty$ limit of the infinite volume Gibbs state.
%{\bf  As $\beta$ tends to infinity, $q_n$ tends to zero for $n\geq 3$, see the discussion following Eq.~(\ref{g(B)}). Therefore, the convergence radius of the activity expansion, the first term in the right member of Eq.~(\ref{A_beta(barmu-beta)}), diverges, but its increase must be compared to that of $\lambda_\beta^d$. Additionally, we also have to take into account the variation of the infinite sum in Eq.~(\ref{A_beta(barmu-beta)}). {\em  I drop the preceding 3 sentences.}}
If $\rho_{v+}(\infty)>0$, there are finite clusters in the ground state at any $\rho<\rho_{\rm cp}\leq\infty$, a possible result for short-range interactions. On the other hand, if $\rho_{v+}(\infty)=0$, there are only infinite clusters in the ground state at arbitrarily low density, a possible result for slowly decaying, but still integrable, interactions.

\section{Summary}

We have proposed a new approach to the problem of classical phase transitions in continuous space. The idea is that  particle motions are coupled via their momenta, and a qualitative change in the coupling occurs as the density increases.
At low density, particles are coupled into finite clusters while at high density, infinite clusters appear, making the transition reminiscent of the percolation transition. To transform the idea into a working method, by expanding the Boltzmann factors into a Fourier series, we have introduced a new set of variables, the momenta of the relative displacements of pairs of particles. Expressed in these variables, the only configurations allowed are those in which the particles are coupled in cyclic clusters. The resulting factorization of particle configurations has led us to a formula for the density that finite clusters can support in the infinite system. In a simplified example, we have demonstrated that in dimensions $d\geq 2$ this density is finite, which implies that at higher densities a positive fraction of the particles must regroup into infinite clusters. Finally, we have shown that in vapor-liquid, vapor-solid and liquid-solid coexistence, the condensed phases appear with infinite clusters, most likely submacroscopic in liquids and macroscopic in crystals.

\vspace{10pt}\noindent
{\bf Acknowledgement.} I thank Mikl\'os Simonovits for helpful discussions on the graph counting problem. This work was supported by the Hungarian Scientific Research Fund (OTKA) through Grant No. K146736.

%\vspace{10pt}\noindent {\bf Conflict of interest:} The author has no conflict of interest or financial interest to declare. The submission is original work and is not under review elsewhere.
%\vspace{10pt}\noindent {\bf Data availability:} No supporting data is available for this article.
%\newpage

\end{document}